\documentclass[aps,pre,twocolumn,floatfix,superscriptaddress,showpacs]{revtex4-1}
\usepackage{graphicx,amsmath,color,amssymb,hyperref}
\usepackage{epstopdf}
\newcommand{\dsh}{d_s^{\textrm{(hub)}}}

\DeclareMathOperator{\for}{for}

\DeclareMathOperator{\asym}{asym}
\DeclareMathOperator{\leaf}{(leaf)}
\DeclareMathOperator{\csch}{csch}
\DeclareMathOperator{\junc}{(junc)}

\begin{document}
\title[]{Origin of the hub spectral dimension in scale-free networks}
\author{S.~\surname{Hwang}}
\affiliation{Department of Physics and Astronomy, Seoul National University, Seoul 151-747, Korea}
\author{D.-S.~\surname{Lee}}
\email{deoksun.lee@inha.ac.kr}
\affiliation{Department of Natural Medical Sciences and Department of Physics, Inha University,
Incheon 402-751, Korea}
\author{B.~\surname{Kahng}}
\email{bkahng@snu.ac.kr}
\affiliation{Department of Physics and Astronomy, Seoul National University, Seoul 151-747, Korea}
\begin{abstract}
The return-to-origin probability and the first passage time
distribution are essential quantities for understanding transport
phenomena in diverse systems. The behaviors of these quantities
typically depend on the spectral dimension $d_s$. However, it was
recently revealed that in scale-free networks these quantities show
a crossover between two power-law regimes characterized by $ d_s $
and the so-called hub spectral dimension $\dsh$ due to the
heterogeneity of connectivities of each node. To understand the
origin of $\dsh$ from a theoretical perspective, we study a random
walk problem on hierarchical scale-free networks by using the
renormalization group (RG) approach. Under the RG transformation,
not only the system size but also the degree of each node changes
due to the scale-free nature of the degree distribution. We show
that the anomalous behavior of random walks involving the hub
spectral dimension $\dsh$ is induced by the conservation of the
power-law degree distribution under the RG transformation.
\end{abstract}


\maketitle

\section{Introduction}
\label{sec:intro} The random walk (RW) approach has been used for
understanding diffusive phenomena in diverse systems
\cite{Hughes1995, Noh2004}. The return-to-origin (RTO) probability and the
first passage time (FPT) distribution are of particular interest and
use to applications. The RTO probability $R_m(t)$ is the probability
that a RWer returns to a starting position $m$ after $t$ time steps
and is determined by the spectral density of the Laplacian matrix
\cite{Rammal1983}. The FPT distribution $F_{s \to m}(t)$ is the
probability that a RWer starting from site $s$ arrives at $m$ at
time $t$ for the first time, and the mean FPT (MFPT) can be obtained
from the first moment of the FPT distribution. Knowledge of the RTO
probability and FPT distribution in diverse systems such as regular
lattices, disordered fractal lattices in Euclidean space, and
complex networks has long been essential in understanding the
transport properties \cite{Avraham2005, Redner2007, Condamin2007, Tejedor2009, Tejedor2011, Zhang2011,
Benichou2010, Reuveni2010, Agliari2009, Meyer2012, Roberts2011}.

It is generally known that the decaying exponents of the RTO
probability and FPT distribution as a function of time are
determined by the spectral dimension $d_s$ \cite{Hughes1995,Kim1984, Hattori1987, Burioni1996,
Haynes2009, Meroz2011, Meyer2011, Samukhin2008, Sood2005}. Moreover,
the MFPT depends on the system size $N$ as $N^{2/d_s}$ for $ d_s <
2$ and as $N$ otherwise \cite{Condamin2007, Haynes2008,
Rozenfeld2007, Zhang2009a, Zhang2009b, Lin2010}. However, a recent study
\cite{Zhang2009} showed that the MFPT scales as $N^{2/d_s}$ even for
the case of a spectral dimension $d_s>2 $ when the target node is a
hub. Moreover, in our previous studies \cite{Hwang2012a,Hwang2012},
we found numerically that the RTO probability and FPT distribution
depend on the degree of the target node in scale-free networks. The
theory we developed in Refs.~\cite{Hwang2012a,Hwang2012} was
successful in explaining these numerical results by introducing the
so-called hub spectral dimension $\dsh$; however, the theory is
based on a heuristic argument and scaling ansatz. Thus, to
illustrate the anomalous behavior of the RTO probability and FPT
distribution due to the presence of the hub, a more rigorous theory
is needed.

To establish rigorous mathematical grounds for our theory, we study
an RW problem on hierarchical scale-free networks
\cite{Hinczewski2006, Rozenfeld2007, Rozenfeld2007a}. We adopt the
renormalization group (RG) transformation approach to trace how the
system is rescaled as the RG transformation proceeds in the
hierarchical network. The occupation probability of the RWers at a
certain node and its rate of change are fundamental quantities for
the RG transformation. The rescaling behavior generates the hub
spectral dimension. The hierarchical network is an ideal toy model
for scale-free networks because it possesses control parameters that
can change the spectral dimension. We calculate the asymptotic
behavior of the RTO probability and FPT distribution as a function
of the control parameters and obtain the hub spectral dimension for
various spectral dimension values.

We also consider the global MFPT (GMFPT), which is the average of
the mean FPT over all starting nodes of RWs to a given destination
node. This quantity can be used for understanding how fast diffusive
spreading takes place on a network. Two types GMFPTs are considered.
In the first type, the average is taken over all starting nodes with
equal weight, while in the second type, the average is taken with a
weight proportional to the source degree. We show that,
counter-intuitively, the qualitative behaviors of the two GMFPTs are
the same.

The paper is organized as follows. In Sec.~\ref{sec:formalism}, we
consider the probability flux of a RWer moving in discrete space and
time and derive the relations between the RTO probability and FPT
distribution for a given starting or target node. We also introduce
the hierarchical model networks. In Sec.~\ref{sec:rto}, we obtain
the RTO probability for the hub starting node by the decimation
method. In Sec.~\ref{sec:fpt}, we derive the FPT distribution and
the MFPT. We then summarize and discuss our findings in
Sec.~\ref{sec:discussion}.

\section{Model and methods}
\label{sec:formalism}

In this section, we first introduce the probability-flux approach for studying the FPT problem \cite{Haynes2008}, 
and explain how to construct the hierarchical networks, called the $(u,v)$-flower model, on which RWs are carried out. 

\subsection{Probability-flux approach}
Let us consider the RWs on a network composed of  $N$ nodes and $L$ undirected links. 
The occupation probability $P_i(t+1)$ of RWs at node $i$ at time $t+1$ satisfies the equation, 
\begin{equation}
P_i(t+1) = \sum_{j \in \mathcal{V}}  \frac{A_{ij}}{k_j} P_j(t),
\label{eq:RW}
\end{equation}
where $\mathcal{V}=\{1,2,\ldots, N\}$ is the set of node indices, $A_{ij}=0$ or $1$, which is the element of the 
adjacency matrix, and $k_j =\sum_{\ell\in \mathcal{V}} A_{j\ell}$ is the degree of node $j$. 
The probability flux from node $i$ to $j$ is defined as
\begin{equation}
Q_{ij}(t) \equiv \frac{P_i(t)}{k_i} - \frac{P_j(t-1)}{k_j}.
\label{eq:FBeqLink}
\end{equation}
This quantity describes the difference between the outgoing flux from $i$ to $j$ at time $t$ and the incoming flux from $j$ 
to $i$ at $t-1$. The net flux from $i$ is then calculated as  
\begin{equation}
Q_i(t) \equiv \sum_{j \in {\rm n.n}(i)}^{}Q_{ij}(t),
\label{eq:FBeq}
\end{equation}
where ${\rm n.n}(i) = \{j\in \mathcal{V}|A_{ij}=1\}$ denotes the set of the nearest neighbor nodes of $i$. If there is no source or sink node in a system, the sum of the occupation probabilities are conserved and the net flux of every node is always zero 
in a steady state. 
From Eqs.~(\ref{eq:FBeqLink}) and (\ref{eq:FBeq}), one finds that the generating functions $q_i(z) \equiv \sum_{t=0}^{\infty}Q_i(t)z^t$ and $p_i(z) \equiv \sum_{t=0}^{\infty}P_i(t)z^t$ are related to each other by 
\begin{eqnarray}
q_i(z) &=& \sum_{j \in \mathcal{V}} \left( k_j \delta_{ij} - z A_{ij} \right) \frac{p_j(z)}{k_j}  \nonumber \\
&\equiv& \sum_{j \in \mathcal{V}} M_{ij}(z) \frac{p_j(z)}{k_j},
\label{eq:FBztrans}
\end{eqnarray}
where we set $P_i(-1) = 0$ for all $i$ and introduced the real symmetric $N\times N$ matrix $M(z)$.
This equation will be solved by the decimation method in the next section under selected initial conditions.   

If the RWer starts at a node $m$, the probability flux becomes 
nonzero at the node $m$ at time $t=0$, as $P_m(0)=1$ and $P_j(-1)=0$ 
for $j$ the neighbor nodes of $m$. Otherwise, the probability flux is zero. 
Therefore, it follows that $Q_i(t)=\delta_{im}\delta_{t0}$ or equivalently
\begin{equation}
q_{i}(z)=\delta_{im}.
\label{eq:bcrto}
\end{equation}
 Under this boundary condition, let us denote the occupation probability at node $i$ as $C_{m\to i}(t)=P_i(t)$. Its generating function $c_{m\to i}(z)$ is then obtained by Eq.~(\ref{eq:FBztrans}) as 
\begin{equation}
c_{m\to i}(z) = k_i M^{-1}_{im}(z),
\label{eq:rtoM}
\end{equation}
which gives the generating function of the RTO probability $r_m(z) = k_m M_{mm}^{-1}(z)$.  
It should be noted that $M(z)$ is symmetric and non-singular if $|z|<1$.

Adopting another boundary condition for the probability flux helps to investigate the FPT distribution of the RW dynamics. Suppose that we study the FPT distribution from node $s$ to $m$. Then the quantity can be computed by investigating the RW with a source of the probability flux at node $s$ and time 
$t=0$ and a sink at node $m$ and time $t=t_{FP}$, where $t_{FP}$ is the time of the first arrival at node $m$ and it can be different for different 
realizations of the RW dynamics.  This condition means that $P_s(0)=1$ and $P_m(t)=0$ for all $t$. 
The ensemble-averaged boundary condition for the probability flux is therefore given by  $Q_{i}(t)=\delta_{is}\delta_{t0}-F_{s\to m}(t)\delta_{im}$ 
with $F_{s\to m} (t)=\langle \delta_{t t_{FP}}\rangle$ being the FPT distribution from node $s$ to $m$. The generating function $q_{i}(z)$ is given as 
\begin{equation}
q_i(z) = \delta_{is} - f_{s\to m}(z) \delta_{im}.
\label{eq:bcfpt}
\end{equation}
For $m\ne s$, the generating function $p_m(z)$ is always zero, i.e., $p_m(z)=0$ since the node is just a sink. Note that $q_m(z)=-f_{s\to m}(z)$.  On the other hand, in case of $m=s$, it holds that $p_m(z)= 1$ since the node is a source, and that $q_m(z) = 1-f_{m\to m}(z)$.  
By Eq.~(\ref{eq:FBztrans}), we find for $m\ne s$ that 
\begin{equation}
0 = p_{m}(z) = k_m\left[M^{-1}_{ms}(z) - f_{s \to m}(z) M^{-1}_{mm}(z)\right]. \nonumber
\end{equation}
For $m=s$, it follows that 
\begin{equation}
1 = p_{m}(z) = k_m \left[ 1 - f_{m\to m}(z)\right] M_{mm}^{-1}(z). \nonumber
\end{equation}
Finally we obtain by using Eq.~(\ref{eq:rtoM})
\begin{equation}
f_{s \to m}(z)  = \frac{k_m}{r_m(z)} \frac{c_{m \to s} - \delta_{ms}}{k_s}
\label{eq:fptrto}
\end{equation}
where we used the relation $M_{ms}^{-1}(z)=M_{sm}^{-1}(z)$.

For the application to the transport from any source node to a given target node, we consider the FPT distribution averaged over all possible starting nodes. One may consider this global FPT (GFPT) distribution of two different types: $F_m(t)$ and $\hat{F}_m(t)$  defined as 
\begin{equation}
F_m(t) \equiv \sum_{s\in \mathcal{V}} \frac{k_s}{2L} F_{s \to m}(t) \nonumber \\
\end{equation}
and
\begin{equation}
\hat{F}_m(t) \equiv \sum_{s \in \mathcal{V}} \frac{1}{N} F_{s \to m} (t).
\label{eq:fmt}
\end{equation}
$F_m(t)$ is the average of $F_{s\to m}(t)$ in the equilibrium state in which the occupation probability is $P_i(t)=k_i/(2L)$. On the other hand, all the starting nodes have equal weights in computing $\hat{F}_m(t)$.
The generating function of $F_m(t)$ is simply related to that of the RTO probability as 
\begin{equation}
f_m(z) = \frac{k_m z}{(2L) (1-z) r_m(z) }.
\label{eq:gfptd}
\end{equation}
The generating function  $\hat{f}_m(z) = \sum_t z^t \hat{F}_m(t)$ involves the occupation probabilities at the nodes other than $m$ as 
\begin{equation}
\hat{f}_m(z) = \frac{k_m}{N r_m(z)} \sum_{s \in \mathcal{V}} \frac{c_{m\to s}(z) - \delta_{sm}}{k_s}.
\label{eq:fptd}
\end{equation}


\subsection{$(u,v)$ flower networks}

\begin{figure}
\includegraphics[width=8.5cm]{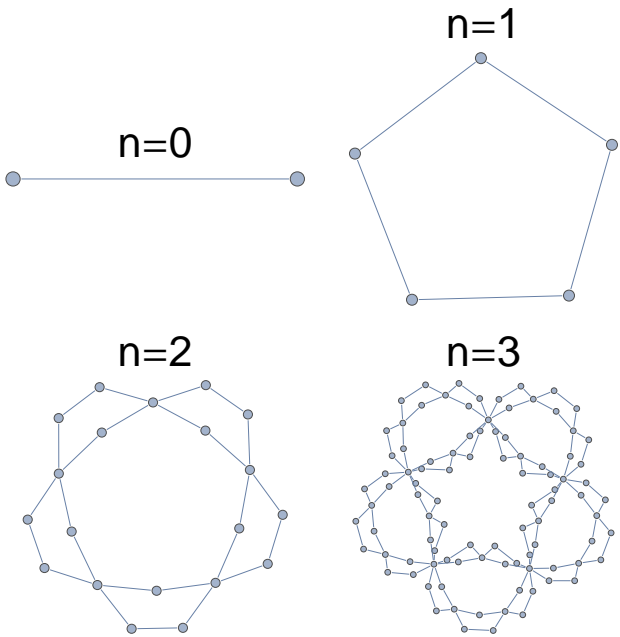}
\caption{Construction of a $(2,3)$ flower network.}
\label{fig:model}
\end{figure}

\begin{table}
\begin{tabular}{lc}
\hline
Number of links& $L^{(n)}= (u+v)^n$ \\
Number of nodes & $N^{(n)}={u+v-2\over u+v-1} (u+v)^n + {u+v\over u+v-1}$\\
Degree of the oldest node & $k^{(n)}=2^n$\\
Degree distribution & $p(k)\sim k^{-\gamma}$ with $\gamma = 1+{\log (u+v)\over \log 2}$\\
Spectral dimension & $d_s = 2 {\log (u+v) \over \log (uv)}$\\
\hline
\label{tab:model}
\end{tabular}
\caption{Basic properties of the $(u,v)$ flower network at generation $n$. }
\label{table:model}
\end{table}

We will derive the RTO probability and the GFPT for hub nodes in the $(u,v)$ flower networks, the construction of which is described  in detail in Refs.~\cite{Hinczewski2006, Rozenfeld2007, Rozenfeld2007a, Hwang2010} and in brief below as well. 
The model network, which we call the {\it flower} network. 
Initially, at generation $0$, there are two nodes connected by a link. The two nodes have indices $0$ and $1$. For $n=1,2,\ldots, $ the network at the $(n+1)$th generation is obtained by replacing each link 
by a \textit{leaf} as described in Fig~\ref{fig:model}.
Each leaf consists of two old nodes at both ends, which existed also in the $n$th generation, and two linear chains consisting of  $u-1$ and $v-1$ new nodes, respectively, 
where $u$ and $v$ are the model parameters.
The growth of the flower network for $u=2$ and $v=3$  is illustrated 
 in Fig.~\ref{fig:model}. 
Basic properties of the flower networks are well 
known~\cite{Hinczewski2006, Rozenfeld2007, Rozenfeld2007a} and also summarized 
in Table~\ref{table:model}.
As noted in Table~\ref{table:model}, the flower networks with different $(u,v)$ may display different characteristic, i.e., being either non-fractal or fractal network.
The spectral dimension may be larger or smaller than $2$ depending on $u$ and $v$.
Therefore, the model network is useful to survey how the spectral dimension influences the behaviors of the RTO probability and the GFPT distribution.
For the later use, we introduce $\mathcal{V}^{(n)}$ to denote the indices of the  nodes present in the $n$th generation of the flower network. We will use the superscript $(n)$  to denote different generations like e.g., $A_{ij}^{(n)}$ and $k_j^{(n)}$.

\section{RTO probability of the hub}
\label{sec:rto}


In this section, we obtain the RTO probability of one of the two oldest nodes, indexed by $0$ and $1$, respectively, in the $(u,v)$ flower network, which are hubs having the largest degree $2^{n}$ at the $n$th generation. 
We below solve Eq.~(\ref{eq:FBztrans}) by performing decimation from the youngest to the oldest nodes to finally obtain $r_0(z)=p_0(z)$ of the oldest (hub) node $0$ in the $(u,v)$ flower network. For the purpose, we introduce the decimation-step-dependent variables $a(s)$ and $\phi_j(s)$'s at each decimation step $s$ with 
\begin{equation}
a(0)= 1/z \ {\rm and} \  \phi_j(0)= {p_i \over  k_i^{(n)} a(0)}
\label{eq:aphi0}
\end{equation}
 to  
rewrite Eq.~(\ref{eq:FBztrans}) as
\begin{equation}
q_i = \sum_{j \in \mathcal{V}^{(n)} } \left(a(0) k_j^{(n)} \delta_{ij} - A_{ij}^{(n)} \right) \phi_j(0).
\label{eq:FBztransUV}
\end{equation}
Here we do not present explicitly the $z$-dependence of $\phi(s)$ and $a(s)$ to focus on their dependencies on the decimation step $s$. 
We are interested in deriving the RTO probability at the $n$th generation, $r_0^{(n)} = p_0 = \phi_0(0) a(0) k_0^{(n)}$.   We first eliminate $\phi_j(0)$ for the youngest nodes, i.e., those with index $j$ in $\mathcal{V}^{(n)} \backslash \mathcal{V}^{(n-1)}$ and rescale the remaining variables 
such that the equations take the same form as in Eq.~\eqref{eq:FBztransUV}.
The decimation process is described in detail in Appendix~\ref{sec:appa}. After  decimating the youngest nodes $s$ times,
the equations for the remaining $\phi$'s read as 
\begin{equation}
q_i = \sum_{j \in \mathcal{V}^{(n-s)}} \left(a(s) k_j^{(n-s)} \delta_{ij} - A_{ij}^{(n-s)} \right) \phi_j(s),
\label{eq:FBztransUV2}
\end{equation}
where $\phi(s)$ and $a(s)$ follow the recursive relations
\begin{equation}
a(s+1) =\frac{\sinh \left((u+v) \cosh ^{-1}(a(s))\right)}{\sinh \left(v \cosh
   ^{-1}(a(s))\right)+\sinh \left(u \cosh ^{-1}(a(s))\right)}
   \label{eq:recursion1}
\end{equation}
and
\begin{equation}
\phi_j(s+1) = \xi(a(s)) \phi_j(s),
\label{eq:recursion2}
\end{equation}
where 
\begin{equation}
\xi(x)= \sqrt{x^2-1}\left[\text{csch}\left(u \cosh ^{-1}x\right)+\text{csch}\left(v \cosh
   ^{-1}x\right)\right].
    \label{eq:xi}
\end{equation}
While one can obtain explicitly the solution of Eq.~(\ref{eq:recursion1}) only for limited cases, i.e., $a(s) = 2^s (a(0) - 1) + 1$ for $(u,v)=(1,2)$ or $a(s) = \cosh \left(u^s \cosh^{-1}a(0)\right)$ when $u=v$, the exact solution to Eq.~(\ref{eq:recursion1}) is not required for evaluating asymptotic behavior of the RTO and the FPT distribution.

We repeat the decimation process until we have 
the equation only for the two initial nodes. The remaining equation is 
\begin{equation}
\left(
\begin{array}{c}
 1 \\
 0
\end{array}
\right) = \left(
\begin{array}{cc}
 a(n) & -1 \\
 -1 & a(n)
\end{array}
\right) \left(
\begin{array}{c}
 \phi_{0}(n) \\
 \phi_{1}(n)
\end{array}
\right),
\label{eq:last}
\end{equation}
and its solution gives $\phi_0(n) = \frac{a(n)}{a(n)^2-1}$. Using $\phi_0(n)$ and Eq.~(\ref{eq:recursion2}), we find that $\phi_0(0)= \phi_0(n) \prod_{s=0}^{n-1} [\xi(a(s))]^{-1}$ and 
\begin{equation}
r_0^{(n)}(z) = \frac{2^n a(n)a(0)}{a(n)^2-1} \prod_{s=0}^{n-1}\frac{1}{ \xi(a(s))},
\label{eq:rto}
\end{equation}
where  $a(0) \equiv 1/z$ and $a(s)$'s for $s\geq 1$ are obtained by Eq.~(\ref{eq:recursion1}). 

To derive the leading singularity of $r_0^{(n)}(z)$,
let us consider  $r_0^{(n-1)}(z^{\prime})$ in the $(n-1)$th generation with  
\begin{eqnarray}
z^{\prime} &=& 1/a(1) \nonumber \\
&=& \frac{\left[\sinh \left(v \cosh^{-1}(z^{-1})\right)+\sinh \left(u \cosh ^{-1}(z^{-1})\right)\right]}{
\left[\sinh \left((u+v) \cosh ^{-1}(z^{-1})\right)\right]}.\nonumber
\end{eqnarray}
Using the same decimation method, we  obtain that 
\begin{eqnarray}
r_0^{(n-1)}(z') &=& \frac{2^{n-1} a(n)a(1)}{a(n)^2-1} \prod_{s=1}^{n-1} \frac{1}{\xi(a(s))} \nonumber \\ 
&=& r_0^{(n)}(z) \frac{a(1)}{2a(0)} \xi(a(0)).
\label{eq:rtoRelation}
\end{eqnarray}
With $\delta \equiv -\ln z$ being small around $z=1$,  $z^{\prime}$ is expanded as $z' \simeq 1- uv \delta + \mathcal{O}(\delta^2) \simeq \exp(- uv \delta)$. Taking the limit $n\to\infty$ with Eq.~(\ref{eq:rtoRelation}), we find that  
\begin{eqnarray}
r_0^{(\infty)}(e^{ - uv \delta}) = r_0^{(\infty)}( e^{-\delta} ) \frac{u+v}{2uv},
\label{eq:rgRelation}
\end{eqnarray}
which, in general, can be satisfied by
\begin{align}
r_0^{(\infty)}(e^{-\delta} ) \sim \delta^{\frac{\ln (u+v)-\ln 2}{\ln(uv) }-1}
\label{eq:r0singularity}
\end{align}
\footnote{Eq.~(\ref{eq:r0singularity}) may have an additional term $A(\delta)$ as $r_0^{(\infty)}(e^{-\delta} ) \sim \delta^{\frac{\ln (u+v)-\ln 2}{\ln(uv) }-1} A(\delta)$, which is the sum of log-periodic function in the form $A(\delta) = \sum_{n=-\infty}^{\infty} A_n |\delta|^{2 \pi i n / \log uv}$.
It is known that such a log-periodic correction term can be negligible~\cite{Domb1977, Grabnera1997}. In the case of $r_0^{(\infty)}(e^{-\delta} )$, the coefficients $A_n$ numerically obtained for (1,2) flower networks~\cite{Grabnera1997} are so small except for $A_0=0.901777$, e.g., as $A_1= 0.00461688 - 0.0251824i$, $A_5 = 0.000198345 - 0.00525088i$, and so on. }.
Here, we used the relation $\xi(1+x) = (u+v) / (uv) + \mathcal{O}(x)$. The singularity of $r^{(\infty)}(z)$ around $z=1$ in 
Eq.~(\ref{eq:r0singularity}) gives the RTO probability $R_m(t)$ via the relation $(d^t/dz^t)(1-z)^\zeta \sim t! t^{-\zeta-1}$ for $t$ large as  
\begin{equation}
R_0^{(\infty)}(t) \simeq c \ t^{-\frac{\ln(u+v)-\ln2}{\ln (uv)}}
=c\  t^{-\frac{\dsh}{2}},
 \label{eq:RTOInfinite}
\end{equation}
where we define the effective spectral dimension of the hub node as
\begin{equation}
\dsh \equiv 2 \frac{\ln \left(\frac{u+v}{2}\right)}{\ln \left(uv \right)}
\end{equation}
The hub spectral dimension $ \dsh $ is related to the spectral dimension $ d_s $ by $ \dsh = d_s (\gamma -2) / (\gamma - 1) $~\cite{Hwang2012a,Hwang2012}.

The overall constant $c$ cannot be determined within our approach. 
The decaying exponent $\dsh$ has been reported in Ref.~\cite{Hwang2012} and the derivation presented here corroborates the result by the exact calculation of the RTO probability in the flower network. 

The generating function of the global RTO probability $ R(t) \equiv \frac{1}{N} \sum_{m=1}^{N} R_m(t) $ satisfies Eq.~\eqref{eq:rgRelation}
with the factor $ 2 $ in the denominator dropped~\cite{Hwang2010}, which gives rise to the difference between $ \dsh $ and $ d_s $ in the $ (u,v) $ flower network.
The factor $ 2 $ arises in Eq.~\eqref{eq:rgRelation} due to the decrease of the degree of the hub node by half every decimation step.
Not only the hub nodes, but also all the surviving nodes do so, and thus the degree distribution preserves its functional form after decimation.

For large but finite $n$, the RTO probability $R_0^{(n)}(t)$ becomes a constant $k_m^{(n)}/(2L^{(n)})$ in the  limit $t\to\infty$, which is preceded by 
the behavior in Eq.~(\ref{eq:RTOInfinite}) as 
\begin{align}
R_0^{(n)}(t) \simeq \left\{
\begin{array}{cc}
c\ t^{-\dsh/2} & \quad  \for \quad 1\ll t \ll  t_c, \\ 
\frac{2^{n}}{(u+v)^n} & \quad \for \quad t \gg t_c,
\end{array} 
\right.
\label{eq:RTOFinite}
\end{align}
\begin{align}
t_c \equiv \left( \frac{c\ (u+v)^n}{2^{n}} \right)^{2\over\dsh} = (2c)^{2\over\dsh} (uv)^n.
\label{eq:tc}
\end{align}
These results are in agreement with Eq.~(16) of Ref.~\cite{Hwang2012a}, considering that the total number of links is given by $ L = (u+v)^n $
Also, the generating function $r_0^{(n)}(z)$ for large but finite $n$ is 
\begin{align}
r_0^{(n)}(e^{-\delta}) \simeq & c\ \delta^{{\dsh \over 2} -1} \gamma\left( 1-\frac{\dsh}{2} , \delta t_c \right) \nonumber \\ &+  \frac{2^{n}}{(u+v)^n} \frac{e^{-\delta t_c}}{\delta},
\label{eq:rtoFinite}
\end{align}
where $\gamma(s,x) = \int_0^{x} t^{s-1} e^{-t} {\rm d} t $ is the lower incomplete gamma function.

\section{GFPT distribution of the hub and its moments}
\label{sec:fpt}

\subsection{Asymptotic behavior of the GFPT distribution of the hub}

We first consider the GFPT distribution of the first type $F_0(t)$ defined in Eq.~(\ref{eq:fmt}) for the hub in the $(u,v)$ flower network. Using Eqs.~\eqref{eq:gfptd} and \eqref{eq:rtoFinite}, one can find the non-analytic behavior of its generation function $f_0^{(n)}(z)$ around $z=1$ at the $n$th generation, which is given by 
\begin{align}
f_0^{(n)}(e^{-\delta}) \simeq \frac{1}{\gamma\left( 1-\frac{\dsh}{2} , \delta t_c \right) \frac{c \sigma^n  \delta^{\dsh\over 2}}{2^{n}} + e^{-\delta t_c }}. 
\label{eq:gfptFinal1}
\end{align}
This result is asymptotically true as $n$ goes to infinity. 
Depending on the magnitude of $\delta t_c$, it shows different behaviors as 
\begin{align}
f_0^{(n)}(e^{-\delta}) \simeq  \left\{
\begin{array}{cc}
 \frac{1}{1+ \frac{\dsh}{2-\dsh} t_c \delta} + \mathcal{O} (\delta t_c)^2   & \quad  \for \quad \delta t_c \ll  1\\ 
\frac{1}{\Gamma(1-\dsh/2)} (t_c \delta)^{-\dsh/2} & \quad  \for \quad \delta t_c \gg  1
\end{array} 
\right.
\label{eq:gfptFinal}
\end{align}
and the GFPT distribution behaves as  
\begin{align}
F_0^{(n)}(t ) \simeq  \left\{
\begin{array}{cc}
 \frac{\sin \left( \pi \dsh/2 \right)}{ \pi t}\left( \frac{t}{t_c} \right)^{\dsh/2}  & \quad  \for \quad t \ll  t_c \\ 
\frac{2-\dsh}{t_c \dsh} \exp(- \frac{(2-\dsh) t}{ \dsh t_c}  ) & \quad  \for \quad t \gg  t_c,
\end{array} 
\right.
\label{eq:GFPTFinal}
\end{align}
where we use the identity $ \Gamma(z) \Gamma(1-z) = \pi / \sin( \pi z) $.
Notice that the GFPT distribution is solely governed by $\dsh$, the hub spectral dimension. 

For the derivation the  GFPT distribution  of the second type $\hat{f}_0^{(n)}(z)$, we refer to the decimation scheme in Sec.~\ref{sec:rto}. We will use  the quantities $a(0)$ and $\phi_j(0)$'s defined in Eq.~(\ref{eq:aphi0}) and 
$a(1)$ and $\phi_j(1)$ in Eqs.~(\ref{eq:recursion1}) and (\ref{eq:recursion2}) with $s=1$  to represent $\hat{f}^{(n)}_0(z)$. To represent explicitly the $z$-dependence of these quantities, we use $\phi_j (z)$ and $\phi_j^{'}(z)$ instead of  $\phi_j(0)$ and $\phi_j(1)$ used in Sec.~\ref{sec:rto}.  Note that $\phi_j^{'}(z)$'s are the renormalized variables of the remaining nodes after decimating out the youngest nodes of 
the $n$th generation. Using the relation $c_{m\to s}(z)/k_s = \phi_s(z) a(0)$, one can represent $\hat{f}^{(n)}_0(z)$ as 
\begin{equation}
\hat{f}_0(z) = \frac{1}{N^{(n)} \phi_0(z)} \sum_{\ell=0}^{n} \phi_{[\ell]}(z) - \frac{1}{N^{(n)} r^{(n)}_0(z)},
\label{eq:fptdUVSetting}
\end{equation}
where $\phi_{[\ell]}(z) \equiv \displaystyle\sum_{s \in \mathcal{V}^{(n)}} \phi_s(z) \delta_{n_s \ell}$ with  $n_s$ is the birth-generation of node $s$.

Summing Eq.~(\ref{eq:FBztransUV2}) over all $i\in \mathcal{V}^{(n-s)}$ with $s=0$ and $s=1$, one can obtain the following normalization relations
\begin{eqnarray}
\sum_{j\in \mathcal{V}} k_j^{(n)}\phi_j(z) &=& {1\over a(0)-1},\nonumber\\
 \sum_{j\in \mathcal{V}^{(n-1)}} k_j^{(n-1)}\phi_j^{'}(z) &=& {1\over a(1)-1},
 \label{eq:norm}
\end{eqnarray}
where we used $\sum_i q_i =1$ from Eq.~(\ref{eq:bcrto}). 
 The degree of node $j$ at the $n$th generation is given by $2^{n-n_s+1}$ except for 
two hubs, the degrees of which are $2^n$.  
Recalling that $\phi_j^{'}(z)=\xi(a(0))\phi_j(z)$ as given in 
Eq.~(\ref{eq:recursion2}),  one can see in Eq.~(\ref{eq:norm}) that the summands of the first line and those  of the second line are identical, if the second line is multiplied by $2/\xi(a(0))$,  for the nodes born at the generations from $0$ to $n-1$.
Therefore, 
we can obtain the values of $\phi_j(z)$'s of the nodes $j$ born at the $n$th generation as 
\begin{equation}
\phi_{[n]}(z) = \frac{1}{2}\frac{1}{a(0)-1} - \frac{1}{\xi(a(0))} \frac{1}{a(1)-1}.
\end{equation}
Extending the relations in Eq.~(\ref{eq:norm}) to the variables in general decimation steps, we can obtain  
\begin{equation}
\phi_{[n-t]}(z) = \frac{1}{2}\frac{\eta_{t}}{a(t)-1} - \frac{\eta_{t+1}}{a(t+1)-1},
\label{eq:generationSum}
\end{equation}
where $\eta_{k} \equiv \displaystyle\prod_{s=0}^{k-1} \frac{1}{\xi(a(s))}$.
Plugging Eq.~\eqref{eq:generationSum} in to Eq.~\eqref{eq:fptdUVSetting}, we  obtain the exact expression for the generating function of the GFPT distribution of the second type given by 
\begin{align}
\hat{f}_0(e^{-\delta}) = & \frac{2^{n-1} a(0)}{N_n r_0(e^{-\delta})} \nonumber \\
&  \left(  \frac{\eta_{0}}{a(0)-1} - \sum_{s=1}^{n-1}  \frac{\eta_{s}}{a(s)-1} - \frac{1}{2^{n-1} a(0)} \right),
\label{eq:fptdUV}
\end{align}
where we inserted the generating function of the RTO probability by using the relation $r_0(z) = \phi_0(z) 2^n a(0)$.
The leading singularity of $\hat{f}_0(z)$ around $z=1$ can be extracted by using the expansions $a(s) = 1 + (uv)^s \delta + \mathcal{O}(\delta^2)$, $\eta_s = (u+v)^s/(uv)^s +  \mathcal{O}(\delta)$ for $\delta = 1-z $ small, which is given in Appendix~\ref{sec:appc},  and $N_n = \frac{ \sigma+(-2+u+v) \sigma^n}{-1+u+v}$ in Table~\ref{table:model}. Then we get,
\begin{equation}
\hat{f}_0(e^{-\delta}) \simeq \frac{2^{n-1}}{(u+v)^{n}} \frac{1}{r_0(e^{-\delta}) \delta} \simeq f_0(e^{-\delta}).
\label{eq:FPTFinal}
\end{equation}
Comparing with $f_0(z)$ of the first type given in Eq.~(\ref{eq:gfptd}), one can see that the GFPT distribution of the second type is exactly the same as that of the first type. 
This implies that the initial distribution of the RWers cannot affect significantly the long time behavior of the GFPT distribution. 

\subsection{Moments of the GFPT distribution}

Here we derive the exact expressions for the moments of the GFPT distribution   in the $(u,v)$ flower network at the $n$th generation. Before presenting the derivation, we consider their scaling properties, which can be identified by the behavior of the generating function $f_0^{(n)}(z)$ in  Eq.~\eqref{eq:gfptFinal}. Since there is no significant difference between the two types of the GFPT distributions, we focus only on the first type $F_m^{(n)}(t)$. 
Its $m$-th moment $\langle t_n^m \rangle = \sum_t t^m F_m (t)$ is obtained from the generating function via  
\begin{equation}
\left\langle t_n^m \right\rangle = \left. (-1)^m \frac{d^m}{d \delta^m} f_0^{(n)}(e^{-\delta}) \right|_{\delta =0},
\label{eq:momentFromGeneratingFunction}
\end{equation}
which yields 
\begin{equation}
\left\langle t_n^m \right\rangle_{\asym} \sim t_c^m m! \sim (uv)^{nm} m! = (N_n^{2/d_s})^m m!.
\label{eq:momentAsymptotic}
\end{equation}
This means that in the $(u,v)$ flower networks at the $n$th generation with finite $n$, $t_c\sim N^{2/d_s}$ acts as the characteristic time scale of the first-passage process, which diverges in the limit $n\to\infty$. Note that it holds even when $d_s > 2$ which resolves the counterexample introduced in Sec.~\ref{sec:intro}.

Beyond the scaling behavior in Eq.~(\ref{eq:momentAsymptotic}), one can derive the exact expression for the first few moments.  
Expanding $a(s)$ and $\xi(a(s))$  for $\delta= -\ln z$ small (see  
Appendix \ref{sec:appc}) and inserting them to Eq.~(\ref{eq:rto}), we find that
\begin{widetext}
\begin{align}
r_0^{(n)}(e^{-\delta}) &= h_{-1} \delta^{-1} + h_0 +  h_1 \delta + \mathcal{O}(\delta^2)\nonumber\\
h_{-1} &=-2^{-1+n} \sigma ^{-n}\nonumber\\
h_0 &= \frac{2^{-2+n} \sigma ^{-n} \left(-2+6 \rho -\sigma ^2+\rho ^n \left(-4+\sigma ^2\right)\right)}{3 (-1+\rho )} \nonumber \\
h_1 &=   \frac{2^{-3+n}  \sigma ^{-n} \left(-4-80 \rho -60 \rho ^2+20 \sigma ^2+20 \rho  \sigma ^2-\sigma ^4-20 \rho ^n (1+\rho ) \left(-4+\sigma ^2\right)+\rho ^{2 n} \left(-16+\sigma ^4\right)\right)}{45 \left(-1+\rho ^2\right)},
\label{eq:rtoSeries}
\end{align}
\end{widetext}
where $\rho=uv$ and $\sigma=u+v$. 
Therefore the GFPT distribution can be obtained by Eq.~\eqref{eq:gfptd} as 
\begin{widetext}
\begin{align}
f_0^{(n)} &= 1-\delta \frac{1+3 \rho -\sigma ^2+\rho ^n \left(-4+\sigma ^2\right)}{6 (-1+\rho )} \nonumber \\ &
+ \delta^2 \left( \frac{9+51 \rho +55 \rho ^2+60 \rho ^3+5 \sigma ^2-60 \rho  \sigma ^2-25 \rho ^2 \sigma ^2+6 \sigma ^4+4 \rho  \sigma ^4}{180 (-1+\rho )^2 (1+\rho )} \right. \nonumber \\ 
& \left. + \frac{5 \rho ^n (1+\rho ) \left(3+5 \rho -2 \sigma ^2\right) \left(-4+\sigma ^2\right)+2 \rho ^{2 n} \left(-4+\sigma ^2\right) \left(-4 (3+2 \rho )+(2+3 \rho ) \sigma ^2\right)}{180 (-1+\rho )^2 (1+\rho )} \right) + \mathcal{O}(\delta^3).
\label{eq:gftpSeries}
\end{align}
\end{widetext}
The mean GFPT is given as 
\begin{eqnarray}
\left\langle t_n \right\rangle &=&  \frac{1+3 \rho -\sigma ^2+\rho ^n \left(-4+\sigma ^2\right)}{6 (-1+\rho )}\nonumber\\
&\simeq&\rho ^n \frac{ \left(-4+\sigma ^2\right)}{6 (-1+\rho )},
\label{eq:t1}
\end{eqnarray}
where the last approximation is valid for large $n$, 
and  again $\rho=uv$ and $\sigma=u+v$.
The second moment $\langle t_n^2\rangle $ for large $n$ is given as
\begin{equation}
\left\langle t_n^2 \right\rangle \simeq \rho ^{2 n} \frac{  \left(-4+\sigma ^2\right) \left(-4 (3+2 \rho )+(2+3 \rho ) \sigma ^2\right)}{90 (-1+\rho )^2 (1+\rho )}.
\label{eq:t2}
\end{equation}

\begin{figure}
\includegraphics[width=8.0cm]{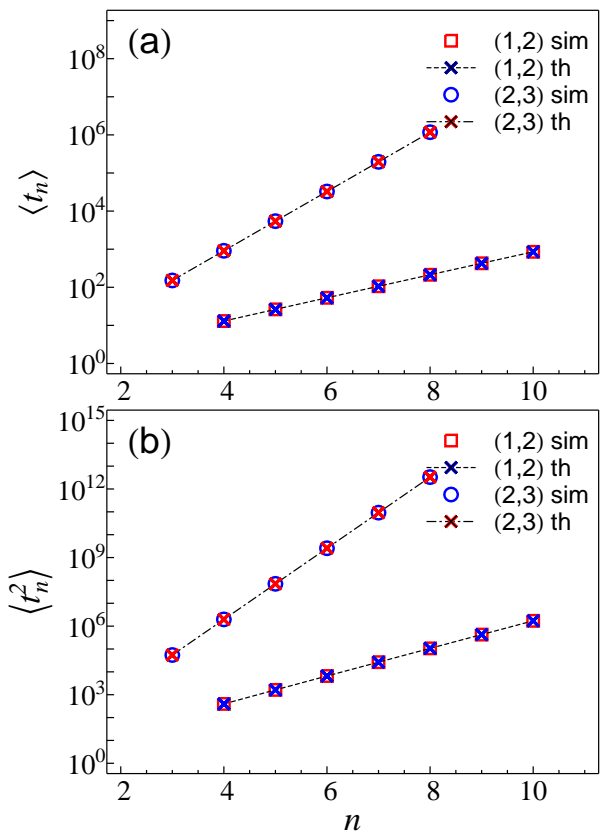}
\caption{(Color online) First two moments of the GFPT distribution for the hub in $(1,2)$ and $(2,3)$ flower networks. 
Dashed lines are the ones obtained theoretically from Eqs.~\eqref{eq:t1} and \eqref{eq:t2}.}
\label{fig:gmfpt}
\end{figure}

Note that first two moments are proportional to $\rho^n$ and $\rho^{2n}$, respectively, as expected in Eq.~\eqref{eq:momentAsymptotic}. 
We can also obtain the exact expressions for the moments of the second-type GFPT distribution $\hat{F}_0^{(n)}(t)$ by inserting Eqs.~\eqref{eq:aSeries}, \eqref{eq:xiSeries} and \eqref{eq:rtoSeries} into Eq.~\eqref{eq:fptdUV} and the results turn out to be identical to Eqs.~(\ref{eq:t1}) and (\ref{eq:t2}) up to a constant in the large-$n$ limit. 
For a numerical check of our results, we performed simulations of RWs on $ (1,2) $ and $ (2,3) $ flower networks to compare the first and second moments of the GFPT distribution with 
Eqs.~\eqref{eq:t1} and \eqref{eq:t2} in Fig.~\ref{fig:gmfpt}.

\section{Summary and Discussion}
\label{sec:discussion}

The main results of this work are given in
Eqs.~\eqref{eq:GFPTFinal}, \eqref{eq:t1}, and \eqref{eq:t2}, in
which the exact GFPT distribution and its moments are presented. 
We formulated the recursive equations relating the fluxes and the
occupation probabilities of the RWs and solved them exactly by using
the decimation method of the RG transformations \cite{Hwang2010}.
Through this method, we were able to derive the exact FPT
distribution for a given node, the hub in the hierarchical network. 
The asymptotic behaviors and the moments of the GFPT and the relation 
between both spectral dimensions are valid for general complex networks~\cite{Hwang2012a, Hwang2012}.
Complementary to the heuristic argument presented in Ref.~\cite{Hwang2012}, 
we derived an exact expression for the
GFPT distribution and its moments. Comparing the RG transformations
of the global and local RTO probabilities, we found that
preservation of the degree distribution during the decimation
process accompanied by the rescaling of node degrees underlies the
appearance of a hub spectral dimension that is different from the
spectral dimension. Considering the practical applications to real
systems, the exact expressions for the mean and variance of the FPT
obtained here will, in addition to the scaling properties obtained
in previous work, be of great use in several problems regarding the
transport phenomena in complex networks.

\begin{acknowledgments}
This work was supported by NRF research
grants funded by MEST (Nos.~2010-0015066 (BK, SH), 2011-0003488 (DSL) and 2011-35B-C00014 (SH)). 
D.-S.L. acknowledges the TJ Park Foundation for support. 
\end{acknowledgments}

\appendix

\section{Derivation of Eqs.~\eqref{eq:recursion1} and \eqref{eq:recursion2} }
\label{sec:appa}
In this Appendix, we derive the recursion relations given in Eqs.~\eqref{eq:recursion1} and \eqref{eq:recursion2} obtained by decimating out the $\phi$ variables from the youngest to the oldest nodes in Eq.~\eqref{eq:FBztransUV}. 
We decompose the $(u,v)$ flower network at the $n$th generation  into the set of leaves, each of which consists of two linear chains of length $u$ and $v$, respectively. The two chains possess $u-1$ and $v-1$ internal (youngest) nodes and share two boundary nodes. Let us consider a chain of length $ (= u$ or $v)$ in a leaf. Then, we use $\phi_i^{\leaf}$ with $i = 1, 2, 3, \cdots, L-1$ to denote the internal nodes on the chain. $\phi_1^{\prime}$ and $\phi_2^{\prime}$ represent two boundary nodes, respectively. 
Then, Eq.~\eqref{eq:FBztransUV} for the internal nodes for each chain can be written as
\begin{eqnarray}
\sum_{l=1}^{L-1} H_{il} \phi_l^{\leaf} = V_i.
\label{eqa:rgleaf}
\end{eqnarray}
with the  $(L-1)\times (L-1)$ matrix $H$ and the column vector $V$ of length $(L-1)$ given by 
\begin{equation}
H=\left(
    \begin{array}{ccccccc}
    2 a & -1 & 0 & \cdots & \cdots & \cdots& 0\\
    -1 & 2 a & -1 & 0 & \cdots & \cdots& 0\\
    0 &-1 & 2 a & -1 & 0 & \cdots& 0\\
    \cdots && \cdots && \cdots && \cdots \\
    0 & \cdots &\cdots & \cdots & 0 & -1 & 2 a
    \end{array}
    \right)
\label{eq:hamiltonian}
\end{equation}
and 
\begin{equation}
V=\left(\phi_1', 0,0, \cdots, 0,\phi_2' \right)^{T}.
\end{equation}
Here the superscript $T$ means the transpose operation.
One can solve Eq.~(\ref{eqa:rgleaf}) to represent all  $\phi_i^{\leaf}$ variables in terms of $\phi_1^{\prime}$ and $\phi_2^{\prime}$.  
To do so, let us consider the eigenvalues $\lambda_\ell$ and the corresponding eigenvectors ${\bf e}_\ell$ with  $\ell=1,2,\ldots,n-1$ of $H$, which  are given by 
~\cite{Cheng2003}
\begin{eqnarray}
\lambda_\ell(L) &=&2 a - 2\cos\left({\ell\pi\over L}\right),\nonumber\\
{\bf e}_\ell(L)&=&\sqrt{2\over L}\left(
    \sin\left({\ell\pi\over L}\right),
    \sin\left({2\ell\pi\over L}\right),
    \ldots\right.\nonumber\\
&& \left.
\ldots,\sin\left({(L-1)\ell\pi\over L}\right)
\right).
\label{eq:lambda}
\end{eqnarray}
Then the matrix $H$ is diagonalized with the bases
\begin{eqnarray}
\tilde{\phi}_\ell^{\leaf} &=& \sqrt{2\over L} \sum_{s=1}^{L-1} \sin\left({\ell s\pi\over L}\right)\phi_s^{\leaf}
\end{eqnarray}
and the solution of Eq.~(\ref{eqa:rgleaf}) in these bases is represented as 
\begin{equation}
\tilde{\phi}_\ell^{\leaf} =  \sqrt{2\over L} \frac{1}{\lambda_\ell (L)} \left( \sin\left(\frac{\ell \pi}{L} \right)  \phi_1' + (-1)^{\ell+1} \sin\left(\frac{\ell \pi}{L} \right)  \phi_2' \right).
\label{eqa:sol}
\end{equation}
Among the $\phi_i^{\leaf}$ variables,  it is only $\phi_1^{\prime}$ and $\phi_2^{\prime}$ that are coupled with $\phi_1^{\leaf}$ and $\phi_{L-1}^{\leaf}$, which are represented as
\begin{align}
\phi_1^{\leaf} =& \frac{2}{L} \sum_{s=1}^{L-1} \frac{ \sin^2 \left( \frac{s \pi}{L} \right)}{\lambda_s(L)} \left( \phi_1^{\prime} + (-1)^{s+1} \phi_2^{\prime} \right)
\label{eq:phi1}
\end{align}
and
\begin{align}
\phi_{L-1}^{\leaf} =& \frac{2}{L} \sum_{s=1}^{L-1} \frac{\sin^2 \left( \frac{s \pi}{L} \right)}{\lambda_s(L)}  \left( \phi_2^{\prime} + (-1)^{s+1} \phi_1^{\prime} \right).
\label{eq:phi2}
\end{align}
When we insert  Eqs.~\eqref{eq:phi1} and \eqref{eq:phi2} into Eq.~\eqref{eq:FBztransUV}, one finds that
\begin{equation}
q_i = \sum_{j \in \mathcal{V}^{(n-1)} } \left(a' k_j^{(n-1)} \delta_{ij} - A_{ij}^{(n-1)} \right) \phi_j^{\prime},
\label{eq:FBztransUVRecursive}
\end{equation}
where $a' \equiv  G/\bar{G}$ and  $\phi_j' \equiv  \bar{G} \phi_j $ with 
\begin{align}
G \equiv 2 a - \frac{2}{u} \sum_{s=1}^{u-1} \frac{\sin^2 \left( \frac{s \pi}{u} \right)}{\lambda_s(u)}-\frac{2}{v} \sum_{s=1}^{v-1} \frac{\sin^2 \left( \frac{s \pi}{v} \right)}{\lambda_s(v)},
\end{align}
\begin{align}
\bar{G}\equiv  \frac{2}{u} \sum_{s=1}^{u-1}  \frac{(-1)^{s+1} \sin^2 \left( \frac{s \pi}{u} \right)}{\lambda_s(u)}   + \frac{2}{v} \sum_{s=1}^{v-1}  \frac{ (-1)^{s+1} \sin^2 \left( \frac{s \pi}{v} \right)}{\lambda_s(v)}.
\end{align}
Note that Eq.~\eqref{eq:FBztransUVRecursive} is in the same form as Eq.~\eqref{eq:FBztransUV} for the $(n-1)$-th generation with the renormalized values $a^{\prime}$ and $\phi_j^{\prime}$. The final recursive relations given in  Eqs.~\eqref{eq:recursion1} and \eqref{eq:recursion2} are obtained by applying the following identities:
\begin{align}
a-\frac{2}{L} \sum_{s=1}^{L-1} \frac{\sin^2 \left( \frac{s \pi}{L} \right)}{\lambda_s(L)} =  \sqrt{a^2-1} \coth( L \cosh^{-1} (a)),
\label{eq:identity1}
\end{align}
\begin{align}
 \frac{2}{L} \sum_{s=1}^{L-1} \frac{ (-1)^{s+1}\sin^2 \left( \frac{s \pi}{L} \right)}{\lambda_s(L)} =  \sqrt{a^2-1} \csch( L \cosh^{-1} (a)),
\label{eq:identity2}
\end{align}
for $a>1$ and $\sinh(x+y) = \sinh x \cosh y + \cosh x \sinh y$. The two identities  in Eqs.~\eqref{eq:identity1} and \eqref{eq:identity2} are  proved below by considering a more general diffusion problem. 

Following the procedures introduced in Ref.~\cite{Haynes2008}, we consider the following diffusion equation for the particle concentration $C(x,t)$ in a one-dimensional continuous space of size $L$ and continuous time: 
\begin{align}
\frac{\partial^2 }{\partial x^2} C(x,t) = \frac{\partial }{\partial t} C(x,t)
\label{eq:diffusion}
\end{align}
with the boundary condition $C(0,t)=\hat{P}_1(t)$, $C(L,t)=\hat{P}_2(t)$, and $C(x,0)=0$. 
Notice that $\hat{P}_1(t)$ and $\hat{P}_2(t)$ are the concentration at both boundaries $x=0$ and $x=L$. 
The concentration-flux at each boundary can be defined by  
$\hat{F}_1(t) = - \left. \frac{\partial}{\partial x} C(x,t) \right|_{x=0}$ and $\hat{F}_2(t) = \left. \frac{\partial}{\partial x} C(x,t) \right|_{x=L}$. Taking the Laplace transformation of Eq.~\eqref{eq:diffusion}, we obtain
\begin{align}
\left( \begin{array}{c}
\hat{f}_1 \\ 
\hat{f}_2
\end{array} \right) = 
\left( \begin{array}{cc}
\alpha' & -\beta' \\ 
-\beta' & \alpha'
\end{array}  \right) \left( \begin{array}{c}
\hat{p}_1 \\ 
\hat{p}_2
\end{array}  \right),
\label{eq:diffusion1}
\end{align}
where $\hat{p}_i(s) = \int_{0}^{\infty} e^{-st}\hat{P}_i(t)  {\rm d} t$ and $\hat{f}_i(s) = \int_{0}^{\infty} e^{-st} \hat{F}_i(t)  {\rm d} t$.
Also we denote $\alpha^{\prime} \equiv \sqrt{s} \coth( L \sqrt{s} )$ and $\beta^{\prime} \equiv \sqrt{s} \csch( L \sqrt{s} )$. 
Now we impose the condition $\hat{f}_1=1$ and $\hat{f}_2=0$ so that the instantaneous source is released at the left boundary at time $t=0$. 

This problem can be analyzed differently by considering the one-dimensional space as a subsequent series of $L$ one-dimensional spaces each of unit size. Then we can relate the Laplace transform of the  flux and the boundary concentrations $\hat{f}^{\junc}_i$'s and $\hat{p}^{\junc}_i$ for $i=1,2,\ldots, L-1$ from 
Eq.~\eqref{eq:diffusion} 
as 
\begin{align}
\left( \begin{array}{c}
\hat{f}_1 \\ 
\hat{f}^{\junc}_{1} \\ 
\hat{f}^{\junc}_{1} \\ 
\vdots \\ 
\hat{f}^{\junc}_{L-1} \\ 
\hat{f}_2
\end{array}  \right) =
 \left( \begin{array}{cccccc}
\alpha & -\beta & 0 & \cdots & 0 & 0 \\ 
-\beta & 2\alpha & -\beta & \cdots & 0 & 0 \\ 
0 & -\beta & 2\alpha & -\beta & 0 & 0 \\ 
\vdots & \vdots & \vdots & \ddots & \vdots & \vdots \\ 
0 & 0 & 0 & -\beta & 2\alpha & -\beta \\ 
0 & 0 & 0 & \cdots & -\beta & \alpha
\end{array}  \right) 
\left( \begin{array}{c}
\hat{p}_1 \\ 
\hat{p}^{\junc}_1 \\ 
\hat{p}^{\junc}_2 \\ 
\vdots \\ 
\hat{p}^{\junc}_{L-1} \\ 
\hat{p}_1
\end{array}  \right).
\label{eq:diffusion2}
\end{align}
Here $\alpha \equiv \sqrt{s} \coth( \sqrt{s} )$ and $\beta \equiv \sqrt{s} \csch( \sqrt{s} )$. 

We can identify Eqs.~\eqref{eq:diffusion2} and \eqref{eq:hamiltonian} by imposing the condition 
$\hat{f}_1=1$, $\hat{f}_2=0$ and $\hat{f}^{\junc}_i=0$ for $i=1,2,\cdots, L-1$ so that 
there is no injection or absorption at each junction.
To make the non-diagonal term  be unity, we divide by $\beta$ Eq.~\eqref{eq:diffusion2}, and set  
$ a = \alpha/\beta$.
Then,  Eq.~\eqref{eq:diffusion2} can be decomposed by using Eq.~\eqref{eq:lambda} similarly 
to Eq.~(\ref{eqa:rgleaf}) and we can obtain
\begin{align}
\hat{p}_1^{\junc} =& \frac{2}{L} \sum_{s=1}^{L-1} \frac{ \sin^2 \left( \frac{s \pi}{L} \right)}{\lambda_s(L)} \left( \hat{p}_1 + (-1)^{s+1} \hat{p}_2 \right)
\label{eq:junc1}
\end{align}
and
\begin{align}
\hat{p}_{L-1}^{\junc} =& \frac{2}{L} \sum_{s=1}^{L-1} \frac{\sin^2 \left( \frac{s \pi}{L} \right)}{\lambda_s(L)}  \left( \hat{p}_2 + (-1)^{s+1} \hat{p}_1 \right).
\label{eq:junc2}
\end{align}
Eliminating $\hat{p}^{\junc}_1$ and $\hat{p}^{\junc}_2$ in Eq.~(\ref{eq:diffusion2}), we  obtain 
\begin{widetext}
\begin{align}
\left( \begin{array}{c}
\frac{1}{\beta} \\ 	
0
\end{array}  \right) =
\left(
\begin{array}{cc}
a-\frac{2}{L} \sum_{s=1}^{L-1} \frac{ \sin^{2} \left( \frac{s \pi}{L} \right) } {\lambda_s(L)} &- \frac{2}{L} \sum_{s=1}^{L-1} \frac{(-1)^{s+1} \sin^{2} \left( \frac{s \pi}{L} \right) } {\lambda_s(L)} \\ 
- \frac{2}{L} \sum_{s=1}^{L-1} \frac{(-1)^{s+1}  \sin^{2} \left(\frac{s \pi}{L} \right) } {\lambda_s(L)} & a-\frac{2}{L} \sum_{s=1}^{L-1} \frac{ \sin^{2} \left( \frac{s \pi}{L} \right) } {\lambda_s(L)} 
\end{array}  \right) 
\left( \begin{array}{c}
\hat{p}_1 \\ 	
\hat{p}_2
\end{array}  \right) 
\label{eq:diffusion3}
\end{align}
\end{widetext}
and comparing Eqs.~\eqref{eq:diffusion1} and \eqref{eq:diffusion3}, we obtain  
\begin{align}
a-\frac{2}{L} \sum_{s=1}^{L-1} \frac{ \sin^{2} \left( \frac{s \pi}{L} \right) } {\lambda_s(L)} = & \frac{\alpha'}{\beta} = \frac{ \coth ( L \sqrt s )}{\csch(\sqrt{s})}, \nonumber
\end{align}
\begin{align}
\frac{2}{L} \sum_{s=1}^{L-1} \frac{ (-1)^{s+1} \sin^{2} \left( \frac{s \pi}{L} \right) } {\lambda_s(L)} = &
\frac{\beta'}{\beta} = \frac{ \csch ( L \sqrt s )}{\csch(\sqrt{s})}.
\end{align}
By considering the identity $ a= \frac{\alpha}{\beta} = \frac{\coth \sqrt s}{\csch \sqrt s} = \cosh \sqrt s$, 
we can replace $\sqrt s$ by $\cosh^{-1} a$, which completes the proof of Eqs.~\eqref{eq:identity1} and \eqref{eq:identity2}.

\section{Expansion of $a(s)$ and $\xi(a(s))$ for $\ln z$ small }
\label{sec:appc}

Eq.~\eqref{eq:recursion1} can be rewritten as $b(s+1) = \mathcal{F}(b(s))$, where $b(s) = a(s)-1$ and 
\begin{widetext}
\begin{equation}
\mathcal{F}(x) = \frac{\sinh \left((u+v) \cosh ^{-1}(x+1)\right)}{\sinh \left(v \cosh
   ^{-1}(x+1)\right)+\sinh \left(u \cosh ^{-1}(x+1)\right)} - 1. 
   \label{eq:functional}
\end{equation}
\end{widetext}
This recursion relation can be exactly solved only when $(u,v)=(1,2)$ or $u=v$. For arbitrary $(u,v)$, we can guarantee the existence of the solution of the form $b(s) = G(t^s c)$ \cite{Kulenovic2002}, where $G(z)$ is an analytic function inside a circle around $z=0$, $t = \left. \frac{ {\rm d} }{ {\rm d} x} \mathcal{F}(x)\right|_{x=0} = uv \equiv \rho$, and $c = G^{-1}(b_0)$. Then the function $G(z)$ satisfies 
\begin{align}
G(\rho z) = \mathcal{F}(G(z))
\label{eq:poincare}
\end{align}
with $G'(0)=1$. This theorem requires the conditions $\mathcal{F}(0)=0$ and $\mathcal{F}'(0)>0$, where Eq.~\eqref{eq:functional} holds.

Therefore, we expand Eq.~\eqref{eq:functional} in the increasing order of $x$ for small $x$ and insert the first few leading terms to Eq.~\eqref{eq:poincare} to find 
\begin{widetext}
\begin{align}
G(z) =z+\frac{z^2 \left(-1+5 \rho -\sigma ^2\right)}{6 (-1+\rho )}+\frac{z^3 \left(1-21 \rho +39 \rho ^2+61 \rho ^3+5 \sigma ^2-18 \rho  \sigma ^2-27 \rho ^2 \sigma ^2+2 \sigma ^4+3 \rho  \sigma ^4\right)}{90 (-1+\rho )^2 (1+\rho )}+ \mathcal{O}(z^4),
\end{align}
\begin{align}
G^{-1}(z) = z+\frac{z^2 \left(1-5 \rho +\sigma ^2\right)}{6 (-1+\rho )}+\frac{z^3 \left(4-24 \rho +36 \rho ^2+64 \rho ^3+5 \sigma ^2-22 \rho  \sigma ^2-23 \rho ^2 \sigma ^2+3 \sigma ^4+2 \rho  \sigma ^4\right)}{90 (-1+\rho )^2 (1+\rho )}.
\end{align}
\end{widetext}
These equations allow us to obtain the series expansion for $a(s)$ given by 
\begin{widetext}
\begin{align}
a(s) &= 1+ G( \rho^s G^{-1}(a(0)-1) ) \nonumber\\
&= 1-\delta  \rho ^s +\frac{\delta ^2 \rho ^s \left(-2-2 \rho +\sigma ^2+\rho ^s \left(-1+5 \rho -\sigma ^2\right)\right)}{6 (-1+\rho )} + \mathcal{O}(\delta^3).
\label{eq:aSeries}
\end{align}
\end{widetext}
Using the recursion relation Eq.~(\ref{eq:xi}) for $ \xi(x) $, we can expand $\xi(a(s))$ as
\begin{widetext}
\begin{align}
\xi(a(s)) &= 
\frac{\sigma }{\rho }+\frac{1}{3} \delta  (-1+\rho ) \rho ^{-1+s} \sigma \nonumber \\
&+\frac{1}{90} \delta ^2 \rho ^{-1+s} \left(10 (1+\rho ) \sigma -3 \rho ^s (-1+\rho  (10+7 \rho )) \sigma +\left(-5+\rho ^s (5+7 \rho )\right) \sigma ^3\right) + \mathcal{O}(\delta^3).
\label{eq:xiSeries}
\end{align}
\end{widetext}

\bibliography{submitted} 

\end{document}